# Ultra-hard rhombohedral carbon from crystal chemistry rationale and first principles

Samir F. Matar[1,§,*], Vladimir L. Solozhenko[2]

[1] Lebanese German University (LGU), Sahel-Alma, Jounieh, Lebanon.
https://orcid.org/0000-0001-5419-358X

[2] LSPM–CNRS, Université Sorbonne Paris Nord, 93430 Villetaneuse, France.
https://orcid.org/0000-0002-0881-9761

[§] *Former DR1-CNRS senior researcher at the University of Bordeaux, ICMCB-CNRS, France.*
[*]Corresponding author email: s.matar@lgu.edu.lb and abouliess@gmail.com

**Abstract**

A new ultra-hard rhombohedral carbon $rh$-$C_4$ (or hexagonal $h$-$C_{12}$) is reported as derived from $3R$ graphite through crystal chemistry construction and ground state energy within the density functional theory. An extended hexagonal three-dimensional network of $h$-$C_{12}$ is formed of $C4$ tetrahedra alike in $h$-$C_4$ lonsdaleite (hexagonal diamond). The electronic band structure of $rh$-$C_4$ is characteristic of insulator with $E_{gap}$ = 4 eV similarly to diamond. From the set of elastic constants a larger value of bulk modulus *versus* lonsdaleite, and the largest Vickers hardness ($H_V$) *versus* both forms of diamond were derived.

# 1 Introduction

Beside hexagonal 2*H* graphite ($C_4$) that is composed of two layers in *AB* sequence, there also exists rhombohedral 3*R* graphite (*rh*-$C_2$) which belongs to the *R*-3*m* space group and is characterized by the three-layers sequence *ABC* [1] (Fig. 1a). Both hexagonal and rhombohedral graphite exhibit a triangular planar coordination of $sp^2$ carbon with an angle of 120°.

Hexagonal diamond *h*-$C_4$ called lonsdaleite (Fig. 1b) was announced by Bundy and Kasper back in 1967 [2], but its actual existence as isolated single crystal is controversial, oppositely to well known (natural and synthetic) cubic diamond (Fig. 1c). Nevertheless, the study of lonsdaleite was extensively developed by theorists due to predicted higher mechanical performance versus diamond as quite recently shown with diamond/lonsdaleite alternating multilayer to model extended biphasic structures letting identify mechanically and thermally composite systems superior to diamond [3]. The investigations were carried out based on computations within the quantum density functional theory (DFT) [4, 5] used herein.

Based on crystal chemistry rationale with subsequent ground state energy calculations within DFT, the purpose of this paper is to devise and propose a new rhombohedral 3D ultra-hard stable carbon (*rh*-$C_4$), expressed as *h*-$C_{12}$ featuring an extended stable analogue of diamond with advanced mechanical properties.

# 2 Crystal chemistry construction rationale

The difference between two graphite structures dwells in the Wyckoff atomic positions which are fixed in 2*H* graphite, i.e. with C1 at 2*b* (0,0,1/4) and C2 at 2*c* (1/3,2/3,1/4) positions while in 3*R* graphite there is one carbon site, and C is at the special twofold 2*c* (*x,x,x*) Wyckoff position (or the six-fold position 6*c* (0,0,*z*) in hexagonal axes settings). Then it becomes possible to consider a change of the stoichiometry from $C_2$ to $C_4$ by occupying additional 2*c* positions with two carbon atoms leading *in fine,* i.e. after full geometry optimization establishing the ground state energetically and structurally, to $C_{12}$ stoichiometry in hexagonal setting. Concomitantly a change of carbon hybridization from $C(sp^2)$ to $C(sp^3)$ carbon is expectedly obtained.

A schematic illustration of the transformation mechanism is shown in Fig. 1d with the red arrows representing the displacements of original carbon atoms of 3*R* graphite now labeled C1, towards the introduced C2, and reaching the proposed new structure (on the right). The resulting displacements lead indeed to purely $sp^3$ carbon (C1 and C2) throughout the structure forming a robust solid 3D network of *C*4 tetrahedra. As a matter of fact it was found from the geometry relaxation calculations that the new structure is characterized by $z_{C1} + z_{C2} = ½$, showing that the two positions are symmetrically interdependent. The structure resulting from a geometry relaxation where C1 and C2 are shown with different colors for sake of clear representation is shown in Fig. 1e. Calculations were also done for the other carbon stoichiometries discussed herein (Table 1) to confront the respective atom averaged total energies.

## 3  Computational framework

The search for the ground state structure and energy was carried out using calculations based on the DFT. The plane-wave Vienna Ab initio Simulation Package VASP code [6,7] was used with the projector augmented wave (PAW) method [7,8] for the atomic potentials with all 4 valence states of the light element C. The exchange-correlation (XC) effects within DFT were considered with the generalized gradient approximation (GGA) [9], this scheme was preferred to local density one which is known to be over-binding. The conjugate-gradient algorithm [10] was used in this computational scheme to relax the atoms onto the ground state. The tetrahedron method with Blöchl *et al.* corrections [11] and Methfessel-Paxton scheme [12] was applied for both geometry relaxation and total energy calculations. Brillouin-zone (BZ) integrals were approximated using a special **k**-point sampling of Monkhorst and Pack [13]. The optimization of the structural parameters was performed until the forces on the atoms were less than 0.02 eV/Å and all stress components less than 0.003 eV/Å$^3$. The calculations were converged at an energy cut-off of 500 eV for the plane-wave basis set concerning the **k**-point integration in the Brillouin zone, with a starting mesh of 6×6×6 up to 12×12×12 for best convergence and relaxation to zero strains. In the post-treatment process of the ground state electronic structures, the charge density and the electronic band structures are computed and illustrated. The mechanical properties are inferred from the set of elastic constants, and the quantities derived from them pertaining to hardness.

## 4  Calculations and discussion of the results

*a- Trends of atom resolved total energies*

In a first step the afore mentioned carbon stoichiometries were examined for their total energies from unconstrained parameter-free calculations along with successive self-consistent cycles at an increasing number of **k**-points. The results are given in Table 1. The total energies are atom averaged to enable comparisons between the different carbon structures; the results are given in the last column. As a general trend, the lowest energy is observed for graphite phases: 3*R* and 2*H* with $E_{tot.}$ = -9.23 eV/at. Slightly lower magnitudes are observed for 3D phases, with the most stable energies observed for diamond and presently proposed *rh*-C$_4$ with a significantly low energy of $E_{tot.}$ = -9.09 eV/at. With a calculation accounting for the whole triple cell C$_{12}$ in hexagonal setting, the energy is slightly lowered by 0.01 eV which could be related with the cell extension. Regarding *h*-C$_4$ (lonsdaleite), the energy $E_{tot.}$ = -9.07 eV/at. is higher than those of diamond and *rh*-C$_4$; it constitutes a remarkable result which will be comforted throughout the developments of the paper.

Lastly for further comparison with a similar 3D carbon form, tetragonal $C_4$ proposed in Ref. [14] was submitted to similar calculations. The atom averaged total energy led to higher of $E_{tot.}$ = -8.89 eV/at., i.e. a less cohesive system than the other 3D phases.

b- *Crystal structures parameters*

The structures of the most stable phases are detailed for $3R$ graphite, $h$-$C_4$ (lonsdaleite) and newly proposed $rh$-$C_4$, and the crystal parameters are given in Table 2 using hexagonal setting of $R$-$3m$ space group (N°166). For $3R$ graphite, a good agreement is found between calculated and experimental values with the noticeable difference of d(C-C) = 1.42 Å, signature of the $C(sp^2)$ in the layered 2D structure. Oppositely in $h$-$C_4$ d(C-C) = 1.54 Å, signature of the tetrahedral $C(sp^3)$ of a 3D structure. The latter result is also found for $rh$-$C_4$ ($h$-$C_{12}$) (Fig. 1e) characterized by a unique d(C-C) = 1.55 Å throughout the lattice (Table 2c). It can be suggested that this arises from the interconnectedness of C1 and C2 positions, i.e. with z(C1) + z(C2) = ½. The structure shown in Fig. 1e presents a large similarity of $h$-$C_{12}$ with $h$-$C_4$ lonsdaleite (Fig. 1b) with the particularity of three times larger cell along $c$-axis. Indeed, Figs. 1e and 1f show multiple 2×2×2 cells of the two structures to further highlight the extension of $h$-$C_{12}$.

c- *Electronic properties from the charge density and the band structures*

Charge density:

To further assess the electronic and crystal structure relationship, the charge density projections onto the chemical constituents as situated in the crystal lattice are needed. The charge density output file resulting from the self-consistent calculations is analyzed with VESTA software [15].

Fig. 2a shows the charge density (yellow) volumes in $3R$ graphite interconnecting *C6* in three layers with continuous yellow rings of $sp^2$-like C. Red zone of slices observed upon crossing crystal planes point to high charge density. Oppositely in Fig. 2b the charge density around carbon atoms in diamond shows a perfect $C(sp^3)$ tetrahedral shape. Fig. 2c sketches $h$-$C_4$ (lonsdaleite) beside the middle of $h$-$C_{12}$ in order to emphasize the similarity of the two structures regarding the succession of carbon along the vertical hexagonal $c$-axis, while highlighting the extended $h$-$C_{12}$ structure. Charge density on the left hand side of Fig. 2c exhibits $4C(sp^3)$ atoms, but a better inspection of successive corner sharing tetrahedra along the vertical hexagonal $c$-axis is observed in $C_{12}$ on the right hand side. Also one can observe the trace of red triangular shape projection of the tetrahedra at the basal planes 0 and 1 showing high charge density featured by the red spots.

Electronic band structures:

Focusing on the 3D structures, Fig. 3 shows the electronic band structures computed using experimental (losdaleite and diamond) and calculated $rh$-$C_4$ ($h$-$C_{12}$). The bands develop along the main directions of the first wedge of the respective hexagonal and face centered cubic Brillouin zones. The hexagonal one is shown in Fig. 3c. All three carbon phases: $h$-$C_{12}$, $h$-$C_4$ and diamond possess large band gaps separating the filled valence band (VB) which show the same band width of 22 eV from the empty conduction band (CB). The zero of energy along the $y$-axis is within the gap. The band gap shows similar magnitude for the three carbon forms with ~4 eV signaling insulating electronic behaviors.

In $h$-$C_4$ lonsdaleite the band gap is indirect between two points in the VB and CB, i.e. between with $\Gamma_{VB}$ and $K_{CB}$, but for both diamond and $rh$-$C_4$ ($h$-$C_{12}$) the gap is from $\Gamma_{VB}$ to a CB minimum in between the lines, A-L and $\Gamma$-X, respectively. Also it is interesting to show that the main difference between the band structures of the two hexagonal phases, besides the number of bands which is proportional to the number of atoms with their valence states, with three times more bands in $h$-$C_{12}$ than in $h$-$C_4$, is the length of the vertical $\Gamma$-A line (cf. Fig. 3c) which is much smaller in the former. This is a reciprocal space property of inversely proportional lengths, here the $c$ parameter longer in $h$-$C_{12}$, whereas the other in-plane lines ($a,b$ plane) are similar in both.

d- Mechanical properties from the full elastic tensors

Elastic constants:

The elastic properties were determined by performing finite distortions of the lattice and deriving the elastic constants from the strain-stress relationship. In hexagonal symmetry there are five independent elastic stiffness constants $C_{11}$, $C_{33}$, $C_{44}$, $C_{12}$, and $C_{13}$. Most encountered compounds are polycrystalline where single-crystal grains are randomly oriented, so that on a large scale such materials can be considered as statistically isotropic. They are then completely described by the bulk modulus $B$ and the shear modulus $G$, which may be obtained by averaging the single-crystal elastic constants. The most widely used averaging method of the elastic stiffness constants is the Voigt one [16] based on a uniform strain. The calculated set of elastic constants is given in Table 3.

All $C_{ij}$ values are positive and their combinations: $C_{11} > C_{12}$, $C_{11}C_{33} > C_{13}^2$ and $(C_{11}+C_{12})C_{33} > 2C_{13}^2$ obey the rules pertaining to the mechanical stability of the phase. The bulk ($B_V$) and shear ($G_V$) moduli following Voigt are formulated for the hexagonal system as:

$$B_V = 1/9 \{2(C_{11} + C_{12}) + 4C_{13} + C_{33}\}$$

and

$$G_V = 1/30 \{C_{11} + C_{12} + 2 C_{33} - 4 C_{13} + 12 C_{44} + 6(C_{11} - C_{12})\}$$

The numerical values are given in the last three columns of Table 3. The highest bulk modulus value is obtained for presently proposed $rh$-$C_4$.

The Pugh's $G/B$ ratio [17] is an indicator of brittleness or ductility for $G/B > 0.5$ and $G/B < 0.5$, respectively. All $G/B$ are > 1 indicating high brittleness.

Hardness:

Vickers hardness ($H_V$) was predicted using three contemporary theoretical models of hardness: (i) Mazhnik-Oganov model [18], (ii) Chen-Niu model [19], and (iii) thermodynamic model [20].. The first two models use the elastic properties, while the thermodynamic approach takes into account the crystal structure and thermodynamic properties.

The results are presented in Table 4 and indicate a slightly higher hardness of rhombohedral carbon compared to diamond (both cubic and hexagonal) in the framework of all three models.

The bulk moduli of $rh$-$C_4$ and lonsdaleite estimated from the thermodynamic model ($B_0$) are in good agreement with those obtained from the elastic constants ($B_V$).

Mazhnik-Oganov model was also used for estimation of fracture toughness ($K_{Ic}$) of dense carbon allotropes (Table 4). For hexagonal and cubic diamond $K_{Ic}$-values are very close, while the fracture toughness of the new phase is slightly higher.

## 5- Conclusion

Crystal chemistry rationale of structurally transforming 2D into 3D followed by geometry-optimization calculations within quantum density functional theory of the generated model structures, were applied to propose a new form of 3D carbon, namely $rh$-$C_4$ (space group $R$-$3m$) from energy and energy-related physical quantities as elastic constants, the charge density and the electronic band structures. Compared to diamond and lonsdaleite, $rh$-$C_4$ is shown to possess slightly better mechanical properties such as hardness and fracture toughness. The novelty of an extended carbon 3D structure together with high performances let suggest $rh$-$C_4$ as a potential ultra-hard material [22].

Table 1. Total and atom averaged energies of the different carbon phases considered in this work (*pw* - present work).

| Carbon phase | Space group | $E_{tot.}$ (eV) | E/at. (eV) |
|---|---|---|---|
| $C_4$ (2H graphite) | $P6_3/mmc$ (N°194) | -36.88 | -9.24 |
| $C_2$ (3R graphite) [1] | $R\text{-}3m$ (N°166) | -18.46 | -9.23 |
| $C_4$ (lonsdaleite) | $P6_3/mmc$ (N°194) | -36.28 | -9.07 |
| $C_8$ (diamond) | $Fd\text{-}3m$ (N°227) | -72.72 | -9.09 |
| $C_4$ (*pw*) | $R\text{-}3m$ (N°166) | -36.36 | -9.09 |
| $C_{12}$ (*pw*) | = with hex. axes | -109.14 | -9.10 |
| $C_4$ (tetragonal) [4] | $I4/mmm$ (N°125) | -35.57 | -8.89 |

*Refs.[1] and [4] are only relevant to the source crystal data used to carry out the energy calculations.*

Table 2. Experimental and (calculated) structure parameters

a) $C_2$ 3R graphite, space group $R\text{-}3m$ (N°166) [2]. Parameters in hexagonal setting:
$a_{hex}$ = 2.456 (2.459) Å; $c_{hex}$ = 10.41 (10.8) Å.

| Atom | Wyckoff | x | y | z |
|---|---|---|---|---|
| C | 6c | 0.0 | 0.0 | 0.167 (0.167) |

d(C-C) =1.419 (1.42) Å.

b) $C_4$ lonsdaleite, space group $P6_3/mmc$ (N°194) [3].
$a_{hex}$ = 2.52 (2.51) Å; $c_{hex}$ = 4.12 (4.17) Å (Fig. 1e).

| Atom | Wyckoff | x | y | z |
|---|---|---|---|---|
| C | 4f | 1/3 | 2/3 | 0.062 (0.063) |

d(C-C) =1.545 (1.54) Å.

c) $C_4$, space group $R\text{-}3m$ (N°166) [*pw*] $a_{rh}$= 4.33 Å; $\alpha$ = 33.57°.
Crystal parameters in hexagonal setting, i.e. $C_{12}$ (Fig. 1e): $a_{hex}$= 2.49 Å; $c_{hex}$ 12.21 Å.

| Atom | Wyckoff | x | y | z |
|---|---|---|---|---|
| C1 | 6c | 0 | 0 | 0.188 |
| C2 | 6c | 0 | 0 | 0.312 |

d(C1-C2) = d(C2-C2) = 1.55 Å.

Table 3. Elastic constants $C_{ij}$ (in GPa) of carbon allotropes

|  | $C_{11}$ | $C_{12}$ | $C_{13}$ | $C_{33}$ | $C_{44}$ | $B_V$ | $G_V$ | $G/B$ |
|---|---|---|---|---|---|---|---|---|
| 3R graphite | 1060 | 180 | 8 | 24 | 415 | 103 | 135 | 1.31 |
| Lonsdaleite (*pw*) | 1180 | 94 | 12 | 1298 | 520 | 432 | 521 | 1.28 |
| *rh*-C$_4$ (*pw*) | 1190 | 92 | 60 | 1323 | 524 | 458 | 552 | 1.20 |
| Diamond (exp) | 1080 | 125 |  |  | 577 | 445 |  |  |

Table 4. Mechanical properties of carbon allotropes: Vickers hardness ($H_V$), bulk modulus ($B_0$), shear modulus ($G$), Young's modulus ($E$), Poisson's ratio ($\nu$) and fracture toughness ($K_{Ic}$)

|  | $H_V$ | | | $B$ | | $G$ | $E$ [§] | $\nu$ [§] | $K_{Ic}$ [†] |
|---|---|---|---|---|---|---|---|---|---|
|  | T[*] | MO[†] | CN[‡] | $B_0$[*] | $B_V$ | | | | |
|  | GPa | | | | | | | | MPa·m$^{½}$ |
| *rh*-C$_4$ | 100 | 105 | 97 | 456 | 458 | 552 $^{(G_V)}$ | 1181 | 0.070 | 6.7 |
| Lonsdaleite | 97 | 99 | 94 | 443 | 432 | 521 $^{(G_V)}$ | 1115 | 0.070 | 6.2 |
| Diamond | 98 | 100 | 93 | 445[**] | | 530[**] | 1138 | 0.074 | 6.4 |

[*] Thermodynamic model [20]

[†] Mazhnik-Oganov model [18]

[‡] Chen-Niu model [19]

[§] $E$ and $\nu$ values calculated using isotropic approximation

[**] Ref. 21

# Figures

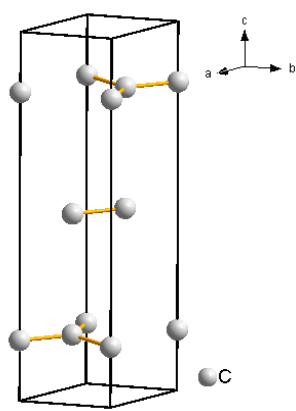 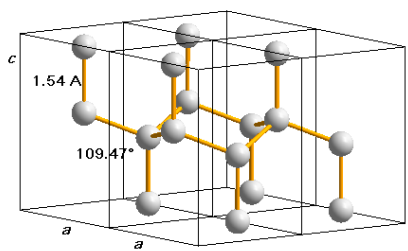 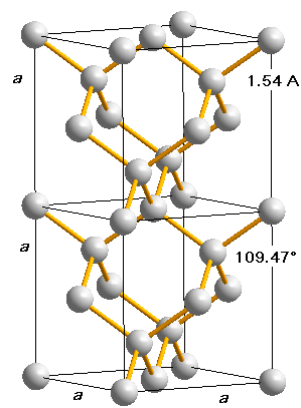

*(a)*        *(b)*        *(c)*

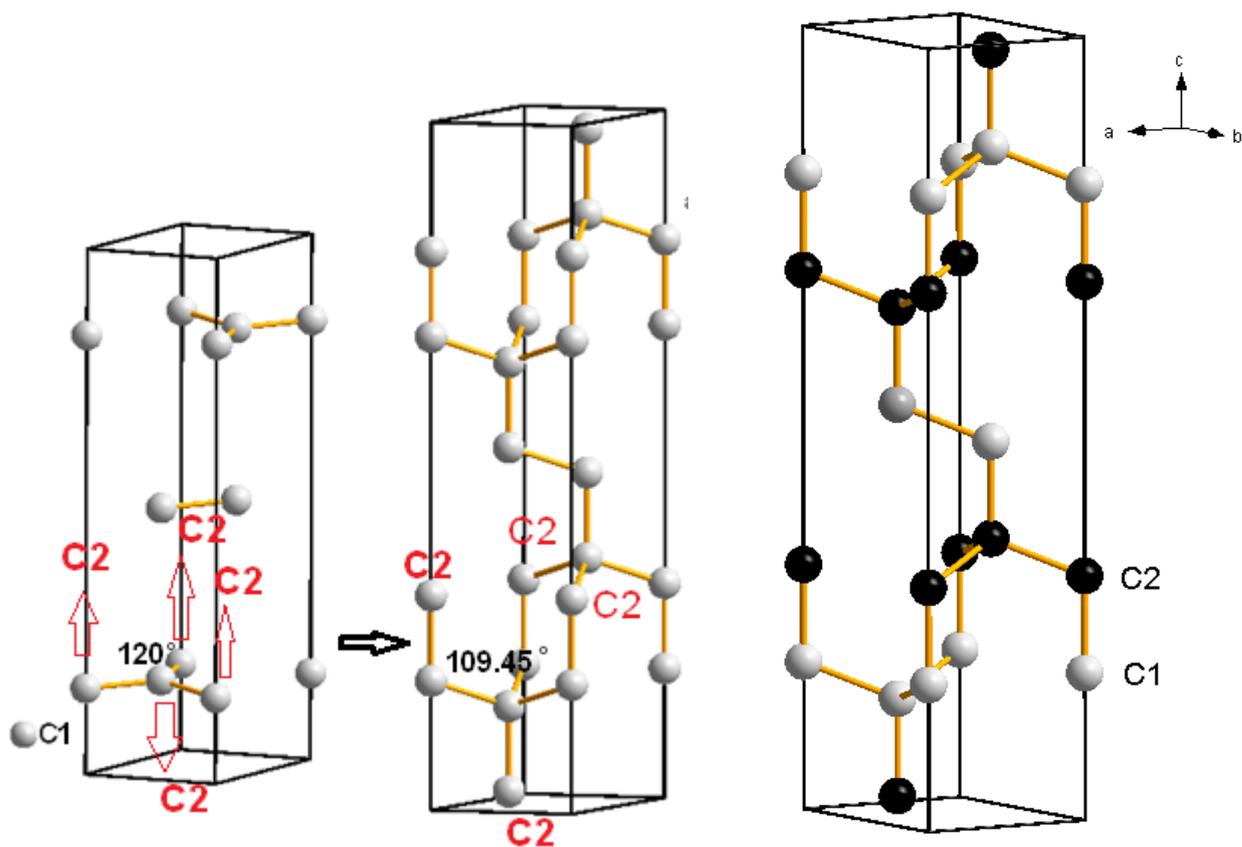

*(d)*        *(e)*

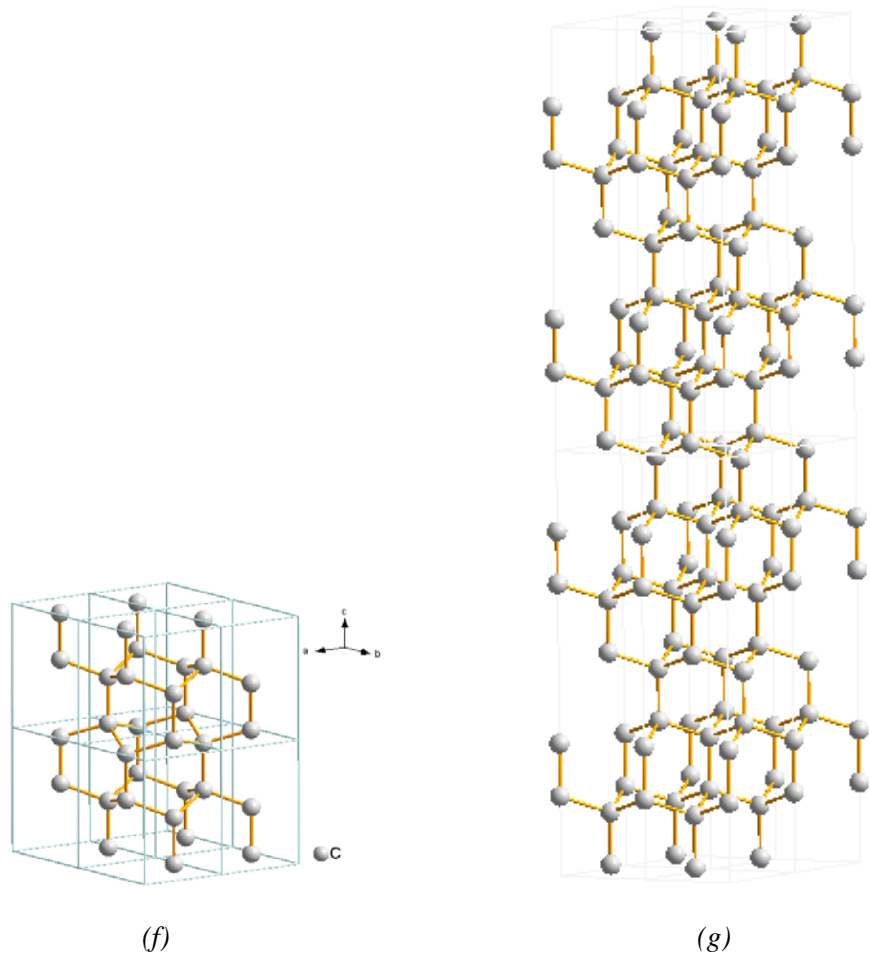

Figure 1. Sketches of the respective structures. (*a*) 3*R* graphite *rh*-C$_2$ (*h*-C$_6$); (*b*) *h*-C$_4$ lonsdaleite 2×2×1 cell; (*c*) *c*-C$_8$ diamond 1×1×2 cell; (*d*) Transformation mechanism 2D → 3D; (*e*) *rh*-C$_4$ (*h*-C$_{12}$) rhombohedral carbon fully geometry opitimized to energy groud state; (*f*) Multiple 2×2×2 cells of lonsdaleite; (*g*) Multiple 2×2×2 cells of C$_{12}$. Rhombohedral structures are represented using hexagonal settings.

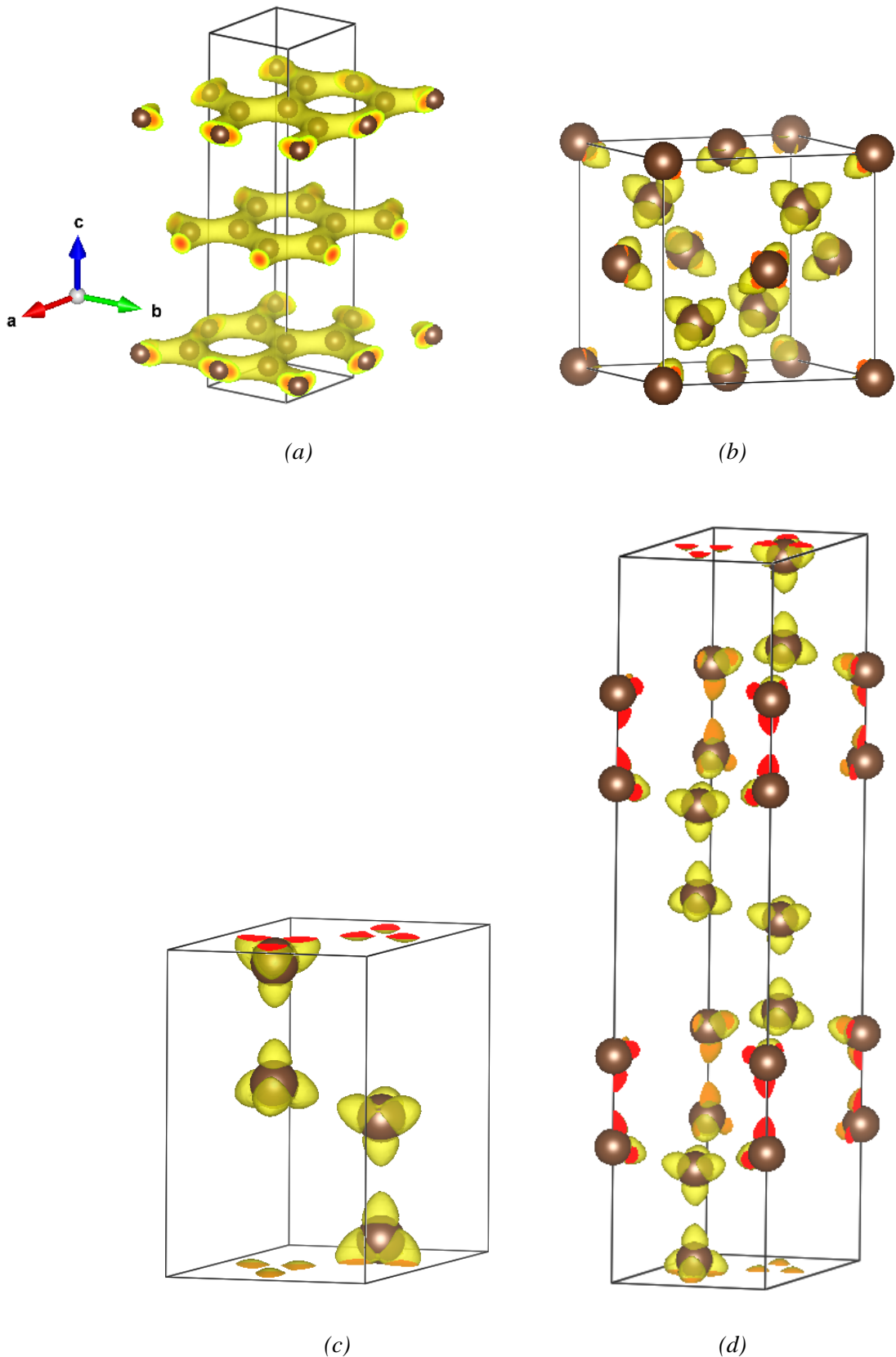

Figure 2. Charge density. C(sp$^2$)-like rings in (*a*) *rh*-C$_2$ 3*R* graphite; and C(sp$^3$)-like tetrahedra in: (*b*) *c*-C$_8$ diamond; (*c*) *h*-C$_4$ lonsdaleite and (*d*) *rh*-C$_4$ (*h*-C$_{12}$) rhombohedral carbon. Brown spheres represent the carbon atoms.

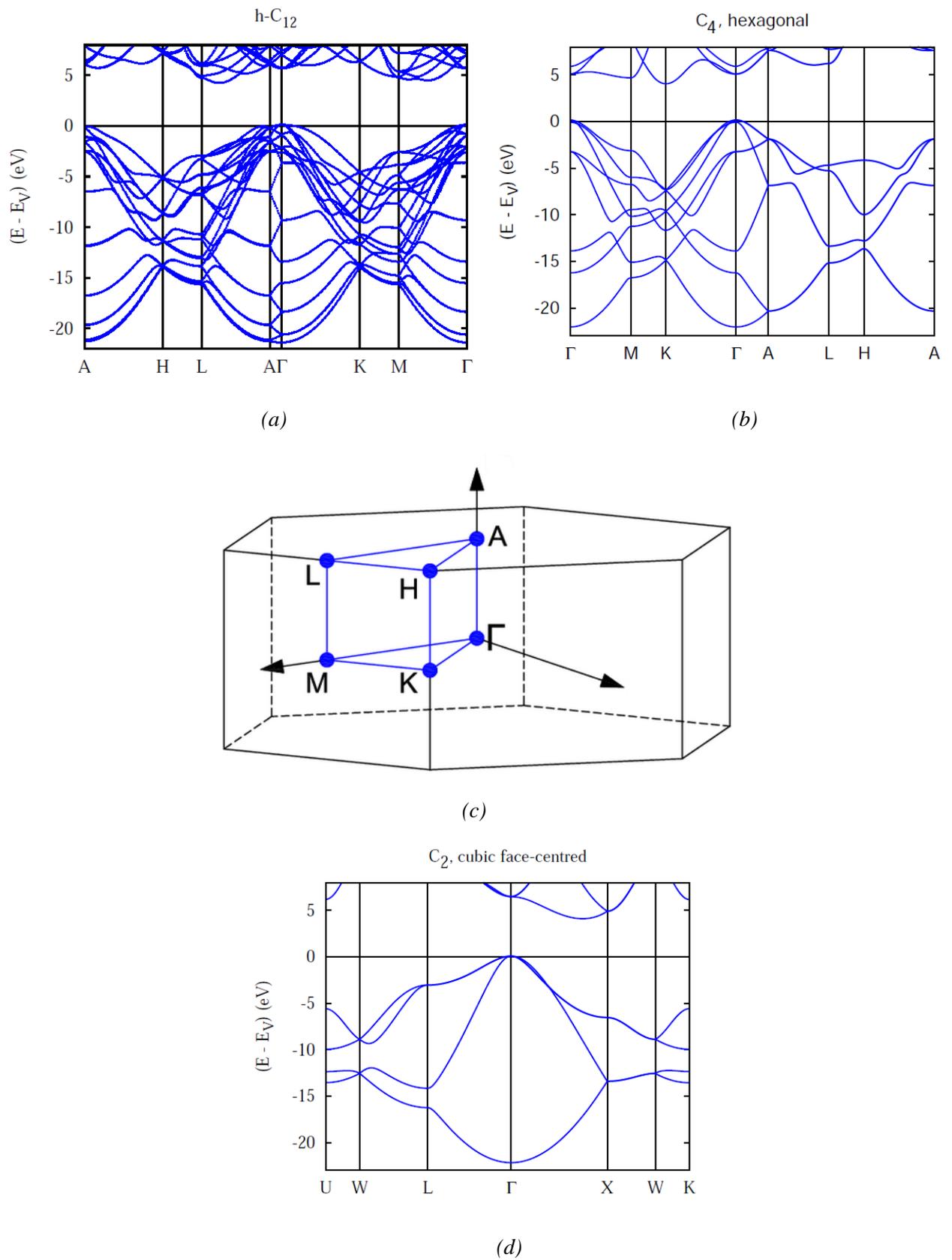

Figure 3. Electronic band structures: (*a*) *rh*-C$_4$ (*h*-C$_{12}$) rhombohedral carbon; (*b*) *h*-C$_4$ lonsdaleite; (*c*) hexagonal Brillouin zone; (*d*) *c*-C$_8$ diamond.